\documentstyle[prl,aps,preprint,tighten]{revtex}
\begin{document}
\draft
\title{Andreev Spectroscopy of Josephson Coupling}
 \author{F.W.J. Hekking$^{(1,2)}$, L. I. Glazman$^{(1)}$, and Gerd
Sch\"on$^{(2)}$}

\address{
 $^{(1)}$ Theoretical Physics Institute, University of Minnesota, \\
Minneapolis, Minnesota 55455, USA\\
$^{(2)}$ Institut f\"ur Theoretische Festk\"orperphysik,
Universit\"at Karlsruhe, \\
 76128 Karlsruhe,  Germany
 }

\date{\today}
\maketitle

\begin{abstract}

It is shown that the low energy spectrum of mesoscopic superconductors
coupled  by Josephson interaction can be probed by two-electron tunneling
from a normal electrode. The Andreev reflection in the NS junction of a
normal-superconductor-superconductor double junction (NSS transistor)
provides a unique spectroscopic tool to probe the
coherent Cooper pair tunneling and the energy
spectrum of the Josephson (SS) junction.  The ground state
properties are reflected in a resonant structure of the linear conductance;
excited states with an energy as low as the Josephson coupling
energy lead to a threshold in the nonlinear $I$-$V$ characteristic.

\end{abstract}

\narrowtext

The quantum states of a mesoscopic conductor (grain)
connected by weak links to particle reservoirs are significantly influenced by
charging effects. The charging energy restricts tunneling of electrons between
the grain and the reservoirs. At a discrete, periodic set of values of the
electrostatic potential of the grain, which can be tuned continuously by means
of a gate, states with different grain charges may have the same electrostatic
energy. In the case of normal metals,
many-body effects lead to a non-analytical variation of the grain charge
with gate voltage in the vicinity of these degeneracy points~\cite{Mat,FSZ}.
However, this behavior develops only at very low temperatures, and has
not been observed  experimentally so far.
The chances for the observation of coherent charge transfer are larger
if the grain and particle reservoirs are in the  superconducting state.
Here the nature of the degeneracy
depends on the relative magnitude of the charging energy $E_C$
and the superconducting gap $\Delta$.
If $\Delta < E_C$ the degeneracy occurs between states having
$N$ and $N+1$ electrons on the grain. In the opposite regime, states with the
numbers $N$ and $N+2$ may be degenerate and coherent Cooper pair
tunneling may become observable~\cite{review}.
The crossover between the two regimes was studied by measurements of the charge
in an electron  box~\cite{Saclay} and of the switching current in a Josephson
double-junction~\cite{Saclay2}.

The possibility of superpositions of different charge
states in a superconductor gives rise to a number of
interesting macroscopic quantum effects~\cite{W,LZ,SZ,Maassen}.
Measurements of the island charge at low temperature and of the critical
Josephson  current provide information about the ground state energy of the
system~\cite{review}, but leave inaccessible the energy spectrum and
structure of excited states\cite{W,LZ,SZ}.
On the other hand, for $\Delta >E_C$, the interplay between the
Josephson coupling and charging leads to quantization of this spectrum. At the
points of charge degeneracy the spacing between the corresponding discrete
levels is of the order of the Josephson coupling energy $E_J$, which for a weak
link is much smaller than the gap $\Delta$. This hierarchy of energies should
allow one to resolve the low-energy spectrum, if a suitable spectroscopic tool
is found.

In this paper we show that the $I$-$V$ characteristic of a double tunnel
junction connecting a superconducting lead with a normal lead through a
mesoscopic superconducting grain, provides the needed
spectroscopic tool. In the low-bias regime, the dominant mechanism of
transport in the NS junction between the grain and the normal electrode is due
to Andreev reflection. Under these conditions we find that
the linear conductance is sensitive to the mixing of different charge states in
the wave functions of the Josephson junction formed by the grain with the
superconducting electrode. In addition, the nonlinear $I$-$V$ characteristic
displays a threshold voltage that allows one to identify the energy of the
excited state.

We consider a NSS transistor with a small superconducting island
coupled via tunnel junctions to a normal left (L) electrode and a
superconducting right (R)  electrode. A gate at potential $V_g$ is coupled
capacitively to the island, and a bias voltage $\pm V/2$ is applied
symmetrically to the electrodes ($V>0$). The charging energy of the
transistor depends on the  number of electrons $N$ on the island, and also on
the number of electrons $N_R-N_L$ which have passed through the transistor,
\begin{equation}
	E_{ch}(N_L,N,N_R) = E_C \left(\frac{Q_g}{e} -N\right)^2
		-\frac{(N_R-N_L)}{2}eV\; .
\label{Ech}
\end{equation}
Here $C = C_L + C_R + C_g$ is the total capacitance of the island, i.e., the
sum of the two junction capacitances and the gate capacitance, and
$E_C=e^2/2C$ is the charging energy corresponding to a single electron. The
effect of the gate voltage is contained in the  external charge
$Q_g = C_g V_g + (C_L -C_R)V$. All the  properties of the transistor are
$2e$-periodic in the external charge. We, therefore, can restrict ourselves
to the range  $0 \le Q_g \le 2e$.

The states of the system $|N_L,N,N_R\rangle$ are characterized by the number
of charges in the electrodes and the island, $N_L, N_R$, and $N$, and
furthermore, if present, by the energies of the excitations in the leads and
the island. We start from a reference state without quasiparticle
excitations, which we denote by $|0,0,0\rangle$ with  charging energy
$E_{ch}(0,0,0) = Q_g^2/2C$.
Other states participating in the two-electron transfer
(Andreev reflection in the left junction and pair tunneling in the
Josephson junction) are
$|0,2,-2\rangle$, $|-2,0,2\rangle$, and $|-2,2,0\rangle$.
Their energies can be found from Eq.~(\ref{Ech}).
The Andreev reflection process involves an intermediate state
$|-1_{\bbox{k}},1_{\bbox{p}},0\rangle$,
where one electron has been transferred through the NS
junction onto the grain.
The energy of this state is $E_{kp} = E_{ch}(-1,1,0) - \xi _k + \epsilon _p$,
where $\xi _k$ and $\epsilon _p$ are the quasiparticle energies in the
normal and the superconducting grain respectively.

The Josephson coupling, which is characterized by the energy scale
$E_J$, mixes charge states differing by multiples of Cooper pairs in
the two superconducting electrodes. If $Q_g$ takes a value near $e$,
and the bias $V$ is small,
the states $|0,0,0\rangle$ and $ |0,2,-2\rangle$ are nearly
degenerate and hence get mixed strongly.
Similarly $|-2,0,2\rangle$ and $ |-2,2,0\rangle$
are mixed. Near the degeneracy points the eigenstates
are a superposition of two states
\begin{eqnarray}
	\psi_0 &=& \alpha|0,0,0\rangle + \beta |0,2,-2\rangle,
\nonumber
\\
	\psi_1 &=& -\beta |0,0,0\rangle + \alpha |0,2,-2\rangle
\label{eigenstates}
\end{eqnarray}
with energies
\begin{equation}
	E_{0(1)} = E_C + \frac{(Q_g -e)^2}{2C} +\frac{eV}{2}
- (+) \frac{1}{2}\sqrt{\delta E_{ch}^2 + E_J^2}
 \; .
\end{equation}
Here the coefficients are
\begin{equation}
	\alpha^2 =  1 - \beta^2 = \frac{1}{2}
	\left[ 1 + \frac{\delta E_{ch}}{\sqrt{\delta E_{ch}^2 +
				E_J^2}}\right]
	\; ,
\label{E01}
\end{equation}
and we introduced the difference in charging energy, $\delta E_{ch} \equiv
E_{ch}(0,2,-2) - E_{ch}(0,0,0)$, which is
$\delta E_{ch} = 4 E_C \left(\frac{Q_g}{e} - 1\right) + eV$.
The coefficient $\alpha$ is close to unity if the charging energy  of the
state $|0,0,0\rangle$  lies below that of $|0,2,-2\rangle$,  i.e. for
$\delta E_{ch} > 0$,  and vanishes in the opposite limit, while $\beta$ shows
the complementary  behavior.  The Josephson mixing of the other two states
leads to the  following eigenstates
\begin{eqnarray}
	\psi'_0 &= \alpha|-2,0,2\rangle &+ \beta |-2,2,0\rangle \; ,
\nonumber
\\
	\psi'_1 &= -\beta |-2,0,2\rangle &+ \alpha |-2,2,0\rangle \; .
\label{eigenstates'}
\end{eqnarray}
The coefficients $\alpha$ and $\beta$ are the same as for the first
pair, and the corresponding energies are
\begin{equation}
	E'_{0(1)} = E_{0(1)} -2eV \; .
\end{equation}
In the above consideration we neglected the effect of charging on
the Josephson coupling  constant that leads to a relatively small
enhancement\cite{SZ,SSS} of  $E_J$.

At low bias voltages the dominant process for charge transfer across the NS
junction is Andreev reflection. Generalizing the expression derived in Ref.
\cite{Hek-Glaz} for a NSN transistor, we can write the amplitude for
this second order tunneling process between the states
$\psi_0$ and $\psi'_0$ as
\begin{eqnarray}
A_{\bbox{k},\bbox{k}'}(\psi_0 \rightarrow \psi'_0)
	&=& \alpha \beta \sum_{\bbox{p}}
	t^*_{\bbox{k}\bbox{p}}t^*_{\bbox{k}'-\bbox{p}}u_p v_p \nonumber \\
 &\times &
	\left(\frac{1}{E_0-E_{\bbox{k}\bbox{p}}}
 +
	\frac{1}{E_0-E_{\bbox{k}'\bbox{p}}} \right)  .
\label{A}
\end{eqnarray}
In this process two electrons from the states
$\bbox{k},\uparrow$ and $\bbox{k}',\downarrow$ of the normal electrode tunnel
into the grain through a junction. The latter is characterized by the
tunneling  Hamiltonian with matrix elements $t_{\bbox{k}\bbox{p}}$. We
suppressed the spin indices, and used the relation
$v_{\bbox{p},\uparrow} = -v_{\bbox{p},\downarrow}$ between the  coefficients
of the Bogoliubov transformation.
The energies of the virtual intermediate states
$E_{\bbox{k}\bbox{p}}$ and $E_{\bbox{k}'\bbox{p}}$ where {\bf one}
electron has been transferred to the island enter the
denominators. The summation in Eq.~(\ref{A}) can be performed
and, for $\xi_k, \xi_{k'} \rightarrow 0$, yields the result
\begin{equation}
	A_{\bbox{k},\bbox{k}'}(\psi_0 \rightarrow \psi'_0)
	=  \alpha \beta \frac{\pi}{2} \nu F_0
	\langle t^*_{\bbox{k}\bbox{p}}t^*_{\bbox{k}'-\bbox{p}}
	\rangle_{_{\hat{\bbox{p}}} } \; .
\label{A'}
\end{equation}
This expression involves the density of states of the island $\nu$, and
 an average over the directions of the
momenta $\hat{p}$.
We introduced the function
\begin{eqnarray}
	F_0 &\equiv & \frac{4}{\pi}
\frac{\Delta}{\sqrt{\Delta^2-[E_{ch}(-1,1,0)-E_0]^2}} \nonumber \\
 &\times &
	\arctan{  \sqrt{\frac{\Delta - E_{ch}(-1,1,0) + E_0}
	{\Delta + E_{ch}(-1,1,0)-E_0} } }  .
\label{F_0}
\end{eqnarray}

The rate for the Andreev reflection process is obtained by the
golden rule. After summation over the initial states $\bbox{k}$
and $\bbox{k}'$ one finds for low temperatures \cite{Hek-Glaz}
\begin{eqnarray}
	&\Gamma &(\psi_0 \rightarrow \psi'_0)\nonumber \\
 &=& \frac{2\pi}{\hbar}
		(\alpha \beta)^2 \frac{(G_n R_K)^2}{16 \pi ^2N_{eff}} F_0^2
		(E_0-E_0')\Theta(E_0-E_0') \; .
\label{gammaA}
\end{eqnarray}
Here $G_n$ is the normal state conductance of the NS junction, $R_K = h/e^2
\approx 25.8 k\Omega$ is the quantum resistance ($G_nR_K \ll 1$), and $N_{eff}$
is the
effective number of  parallel channels in the tunnel
junction\cite{Hek-Glaz,HN}. These parameters are conveniently absorbed in the
definition of the Andreev conductance, $G_A = G_n^2 R_K/N_{eff}$.
If the applied bias $V$ is below some threshold voltage $V_{\rm th}$, the only
transition possible at low temperatures is the Andreev reflection between the
states $\psi_0$ and $\psi'_0$. The resulting current   $I_{\rm res} = 2e
\Gamma(\psi_0 \rightarrow \psi'_0)$, due to the overlap of the functions
$\alpha$ and $\beta$, shows a pronounced structure as a function of
gate charge, typical for resonant tunneling. This is most clearly seen
in the dependence of the linear conductance on the gate charge
\begin{equation}
        G_{\rm res}(Q_g) = G_A \frac{E_J^2}{16 E_{C}^2(Q_g/e - 1)^2+E_J^2}
\frac{F_0^2}{4} \;.
\label{res}
\end{equation}
The result is illustrated by the lower set of curves in Fig.~1.
The width of the resonance is characterized by $E_J$.
On this energy scale,
as long as $\Delta -E_C \agt E_J$,
the function $F_0$ can be considered constant.

Eq. (\ref{res}) was derived under the assumption
that the energy $\Delta + E_{ch}(-1,1,0)$ of the intermediate
state lies above $E_0$.
The resonant shape changes drastically if the superconducting gap is lowered,
such that these two energies can coincide.
This occurs at two values of charge, $Q_g= e\pm \delta Q_g^*$,
where $\delta Q_g^*$ is
\begin{equation}
	\frac{\delta Q_g^*}{e}
	= \frac{1}{2} \sqrt{\left( 1- \frac{\Delta}{E_C}\right)^2
	- \left( \frac{E_J}{2 E_C} \right)^2} \; .
\label{divergence}
\end{equation}
If $Q_g$ lies within the window
$e- \delta Q_g^* \le Q_g \le e + \delta Q_g^*$,
the ground state is no longer composed of even-charge states;
rather the state $|-1,1_{\bbox{p}},0\rangle$ with one electron
transferred through the NS junction has the lowest
energy~\cite{Saclay,Av-Naz}.
In the odd state the current is low;
the maximal conductance $G_{\rm res} (e)$, see Eq.~(\ref{res}),
is not reached.
Instead, within the abovementioned charge window, the conductance
assumes the low value
$G \sim G_A(G_nR_K)^2$ determined by higher order tunneling
processes through both junctions~\cite{footnote}.

Even at $\Delta > E_C - E_J/2$,
 Andreev reflection leads solely to transitions between the states
$\psi _0$ and $\psi ' _0$ only, as long as
 the bias voltages are below a certain threshold.
This threshold voltage is determined by the
condition $V_{\rm th}=(E_1-E_0)/2e$.
Its smallest value  is $V_{\rm th}=E_J/2e$.
If the junction capacitances are equal, $C_L = C_R$, we find
\begin{equation}
eV_{\rm th}
=
\frac{4}{3}E_C \left(\frac{Q_g}{e} -1 \right) +
\sqrt{\left[\frac{8}{3}E_C\left(\frac{Q_g}{e}-1\right)\right]^2
+\frac{1}{3}E_J^2}.
\label{Vth}
\end{equation}
At higher voltages, $V>V_{\rm th}$, Andreev reflection processes can also
lead to transitions between  the other states introduced above.
In addition to the transition $\Gamma(\psi_0 \rightarrow \psi'_0)$ we find
\begin{eqnarray}
\Gamma(\psi_0 \rightarrow \psi'_1) &=& \alpha ^4
\frac{G_A}{4e^2}F_0^2 \,
[2eV-(E_1-E_0)]\Theta(V-V_{\rm th}) \; ,
\nonumber
\\
	\Gamma(\psi_1 \rightarrow \psi'_0) &=& \beta ^4
\frac{G_A}{4e^2} F_1^2 \,[2eV+(E_1-E_0)] \; ,
\nonumber
\\
	\Gamma(\psi_1 \rightarrow \psi'_1) &=& (\alpha \beta)^2
\frac{G_A}{4e^2} F_1^2 \, 2eV \; .
\label{Gammas}
\end{eqnarray}
The function $F_1$ is defined similar
to $F_0$, but with the energy of the initial state $E_0$ replaced by $E_1$.
Both $F_0$ and $F_1$ are approximately constant on the energy scale $E_J$
if $\Delta -E_C \agt E_J$.
A master equation yields the probabilities $W_1$, $W_0$ for the system to be
 in the excited and ground state, respectively. I.e.,
\begin{eqnarray}
	W_1 &=& \frac{\Gamma(\psi_0 \rightarrow \psi'_1)}
	{\Gamma(\psi_0 \rightarrow \psi'_1)
			+ \Gamma(\psi_1 \rightarrow \psi'_0)}\Theta (V-V_{\rm th}), \nonumber \\
W_0&=&1-W_1.
\label{ww}
\end{eqnarray}
The current then is
\begin{eqnarray}
	I &=& 2e [\Gamma(\psi_0 \rightarrow \psi'_0)
			+ \Gamma(\psi_0 \rightarrow \psi'_1)] W_0 \nonumber \\
  &+&
		2e [\Gamma(\psi_1 \rightarrow \psi'_1)
		+ \Gamma(\psi_1 \rightarrow \psi'_0)] W_1  .
\label{I>}
\end{eqnarray}
The threshold dependence of the probability $W_1$
on $V-V_{\rm th}$ allows one to determine the Josephson energy $E_J$.
This is illustrated with the help of Fig.~1.
As long as the applied bias $eV$ is smaller than
$E_J/2$, which is the smallest value of $eV_{\rm th}$,
the gate voltage dependence
of the differential conductance $G=dI/dV$ shows the resonance~(\ref{res})
(two lower curves in Fig.~1). As soon as $eV > E_J/2$, the probability
$W_1$ can become nonzero,
and an additional channel for charge transfer opens up
at two values of $Q_g$ (cf. Eq. (\ref{Vth})). This leads
to a stepwise change of the conductance
(two upper curves in Fig.~1).
The  smallest bias voltage at which these
jumps occur enables one to determine $E_J$.
The magnitude $\delta G _{\pm}$
of the two jumps depends on the applied bias. If $C_L=C_R$
we find
\begin{eqnarray}
&& \delta G _{\pm} (V)
=
 G_A \times
\nonumber \\
 && \left\{1  \pm   \left[1+\frac{1}{2}
\left(\frac{E_J}{2eV}\right)^2\right]
\sqrt{1-\left(\frac{E_J}{2eV}\right)^2}\right\}\frac{F_1^2}{2}.
\label{deltaG}
\end{eqnarray}
Another way to detect the Josephson energy is a measurement
of the differential conductance as a function of
the bias voltage $V$ at a fixed gate-potential, illustrated in Fig. 2.
Depending on the value of $Q_g$, a jump
occurs at a certain threshold $V_{\rm th}$, cf. Eq. (\ref{Vth});
the magnitude $\delta G (V_{\rm th})$
is given by Eq.(\ref{deltaG}).

Spectroscopy of the Josephson coupling is only possible
as long as real single electron tunneling processes are suppressed.
Similar to the case of an NSN junction~\cite{Hek-Glaz},
the Andreev reflection can get ``poisoned'' once such processes
become possible.
Poisoning occurs at a certain threshold bias $V_{\rm poison}$,
at which the energy of the odd-charge state is lower than $E_0$.
 If the applied bias $eV$ exceeds $eV_{\rm poison}$ by an amount as
little as the level spacing of the grain,
the probability $W_o$ for the system to be in an odd charge
state is approximately unity and current will drop substantially.
Therefore, for the spectroscopy of the Josephson coupling, the
condition $V_{\rm th} < V_{\rm poison}$ must be met. At resonance
$Q_g \simeq e$
this condition is satisfied if $\Delta>E_C+E_J/2$.

The resonance structure in the $I(Q_g)$-dependence and the threshold
structure in the $I(V)$-dependence is most pronounced if the temperature
is smaller than the width of these
features, which is given by the Josephson coupling energy $E_J$.
In addition to intrinsic thermal
fluctuations there exist fluctuations due to the external
electrodynamic environment. They persist down to the lowest temperatures,
causing voltage fluctuations proportional to the resistance of the
external circuit. Therefore, a low impedance environment creates the
most favorable conditions for the observation of the effects discussed here.
It is also essential that there are no quasiparticles in the grain.
This requires a sufficiently large
superconducting gap $\Delta > E_C + E_J/2$, a perfect BCS density
of states, and not too large bias voltages. In exchange for these
restrictions we have a well controlled theory and unambiguous
predictions for experiments. We can mention that coherent tunneling
of Cooper pairs plays a role in a number of effects. Some
examples are the gate voltage dependence of the critical current of SSS
transistors~\cite{Saclay2,SSS},
and the resonant Cooper pair tunneling
at finite bias voltage~\cite{Maassen,Havi}.
The latter effect is controlled by quasiparticle-induced dissipation
or influence of the
environment.
The spectroscopy of coherent
mixing by Andreev reflection offers significant advantages over
these examples.
It is more
straightforward than inferring the critical current from the runaway value
\cite{Saclay2},
and requires only a direct measurement of the $I$-$V$
characteristic in a convenient regime of subgap voltages.
In contrast to the experiments of Haviland et al.
\cite{Havi}, the suggested method does not rely on the fluctuations due to
the hardly
controllable electrodynamic environment of the junctions.

This work is supported by ``Sonderforschungsbereich 195'' of the Deutsche
Forschungsgemeinschaft, by NSF Grant No. DMR-9423244, and by the Netherlands
Organization for Scientific Research (NWO). One of the authors
(LG) acknowledges the hospitality of the University of Karlsruhe
where this work has been performed.

\begin{figure}
\caption{Differential conductance $G/G_A$ of a NSS transistor as a function
of the gate charge $Q_g/e$, for various values of $eV/E_C = .047, .049,
.051, .053$ (from bottom to top, curves off-set for clarity).
The parameters are: $E_J/E_C = 0.1$, $C_L = C_R$, and
$\Delta$ is large compared to $E_C$. The two lower curves ($eV<E_J/2$)
correspond to the resonance~(11), the upper curves ($eV > E_J/2$) show how
threshold processes affect this resonance.}
\end{figure}

\begin{figure}
\caption{Differential conductance $G/G_A$ as a function of bias voltage
$eV/E_C$, for various values of $Q_g/e = .99, 1.0, 1.01$
(from bottom to top, curves off-set), parameters as in Fig. 1.
A jump occurs at $V=V_{\rm th}$, Eq.~(14). Its magnitude is given
by Eq.~(18): $\delta G _+$ if $Q_g/e > 1 - eV_{\rm th}/4E_C$,
and $\delta G _-$ in the opposite limit.}
\end{figure}


\begin{references}


\bibitem{Mat}
K.A. Matveev, Sov. Phys. JETP {\bf 72}, 892  (1991); L. I. Glazman,
K.A. Matveev, Sov. Phys. JETP {\bf 71. }, 1031 (1990).

\bibitem{FSZ}
G. Falci, J. Heinz, G. Sch\"on, and G.T. Zimanyi, to be published in
Physica B; K.A. Matveev, ibid; H. Grabert, ibid.;
H. Schoeller and G. Sch\"on, to be published in Phys. Rev. B.

\bibitem{review} For a review see e.g.  K.A. Matveev, L. I. Glazman,
and R.I. Shekter, Mod. Phys. Lett. {\bf B 8}, No 15 (1994).

\bibitem{Saclay} P. Lafarge, P. Joyez, D. Esteve, C. Urbina, and M.H. Devoret,
 Phys. Rev. Lett. {\bf 70}, 994 (1993)

\bibitem{Saclay2} P. Joyez, P. Lafarge, A. Filipe, D. Esteve, and M.H. Devoret,
 Phys. Rev. Lett. {\bf 72}, 2458 (1994)

\bibitem{W}
A. Widom, G. Megaloudis, T.D. Clark, H. Prance, and R.J. Prance,
J. Phys. A {\bf 15} 3877 (1982).

\bibitem{LZ}
K.K. Likharev and A.B. Zorin, J. Low Temp. Phys. {\bf 59} 347 (1985);
D.V. Averin and K.K. Likharev, in {\it Mesoscopic Phenomena in Solids},
B.L. Altshuler et al. eds., p. 173 (Elsevier,1991).

\bibitem{SZ}
G. Sch\"on and A.D. Zaikin, Phys. Rep. {\bf198}, 237 (1990).

\bibitem{Maassen} A. Maassen v.d. Brink, G. Sch\"on, and
L.J. Geerligs, Phys. Rev. Lett. {\bf 67}, 3030 (1991);
A. Maassen v.d. Brink et al., Z. Phys. B {\bf 85}, 459 (1991)

\bibitem{SSS} K.A. Matveev, M. Gisself\"alt, L. I. Glazman, M. Jonson,
and R.I. Shekter, Phys. Rev. Lett. {\bf 70}, 2940 (1993).

\bibitem{Hek-Glaz} F.W.J. Hekking, L.I. Glazman, K.A. Matveev, and
R.I. Shekhter, Phys. Rev. Lett. {\bf 70}, 4138 (1993).

\bibitem{HN} F.W.J. Hekking and Yu.V. Nazarov, Phys. Rev. Lett. {\bf 71}, 1625
(1993).

\bibitem{Av-Naz} D.V. Averin and Yu.V. Nazarov, Phys. Rev. Lett. {\bf
69}, 1993 (1992).

\bibitem{footnote}We note, though, that it is larger than the conductance
in the odd state of an
NSN transistor~\cite{Hek-Glaz} with the same parameters of the tunnel
junctions by a
factor $N_{eff}$, which is typically of the order of $50$.

\bibitem{Havi} D.B. Haviland et al., Phys. Rev. Lett. {\bf 73}, 1541 (1994)
\end{references}
\end{document}